\documentclass[journal=jacsat,manuscript=article]{achemso}

\usepackage{chemformula} 
\usepackage[T1]{fontenc} 
\usepackage{color}
\usepackage{graphicx}
\usepackage{dcolumn}
\usepackage{bm}
\usepackage{comment}
\usepackage[english]{babel}
\usepackage[latin1]{inputenc}
\usepackage{siunitx}
\usepackage{hyperref}
\DeclareSIUnit\angstrom{\text {\AA}}
\DeclareSIUnit\bar{\text bar}

\newcommand{\SIadj}[2]{\SI[number-unit-product={\text{-}}]{#1}{#2}}
\author{Jingtian Zhao}
\affiliation[Unknown University]
{Zernike Institute for Advanced Materials, University of Groningen, 9747AG Groningen, The Netherlands.}
\alsoaffiliation{Groningen Cognitive Systems and Materials Center (CogniGron), University of Groningen, 9747AG Groningen, The Netherlands.}

\author{Beatriz Noheda}
\affiliation[Unknown University]
{Zernike Institute for Advanced Materials, University of Groningen, 9747AG Groningen, The Netherlands.}
\alsoaffiliation{Groningen Cognitive Systems and Materials Center (CogniGron), University of Groningen, 9747AG Groningen, The Netherlands.}

\author{Martin F.\ Sarott}
\email{m.f.sarott@rug.nl}
\affiliation[Unknown University]
{Zernike Institute for Advanced Materials, University of Groningen, 9747AG Groningen, The Netherlands.}
\alsoaffiliation{Groningen Cognitive Systems and Materials Center (CogniGron), University of Groningen, 9747AG Groningen, The Netherlands.}

\title{Integration of imprint-free and low coercivity ferroelectric BaTiO\textsubscript{3} thin films on silicon}


\begin{document}







\begin{abstract}
Highly-crystalline ferroelectric oxides integrated on Si hold great promise for energy-efficient memory and logic technologies. 
Exploiting epitaxial strain engineering in these materials is, however, severely hampered on Si, where the large structural mismatch often results in an inferior interfacial quality and causes a degradation of the ferroelectric switching characteristics.
In this work, we present the growth of single-crystalline BaTiO\textsubscript{3} thin films on Si, exhibiting imprint-free switching, low coercivity, high remanent polarization, and no fatigue for over $10^{10}$ switching cycles. 
We accomplish this via the insertion of a SrSn\textsubscript{1-x}Ti\textsubscript{x}O\textsubscript{3} layer on SrTiO\textsubscript{3}-buffered Si.
This layer serves as a pseudo substrate that alleviates the thermal strain that the Si substrates imposes on the BaTiO\textsubscript{3} layer, while simultaneously providing moderate compressive strain that stabilizes a pure out-of-plane polarization.
Thus, our work paves the way toward the fabrication of Si-compatible, low-power-consuming ferroelectric devices for non-volatile memory applications.
\end{abstract}

\section{Introduction}

The growing demand for energy-efficient memory and logic devices drives the need to integrate new materials with additional functional properties into the complementary metal-oxide semiconductor (CMOS) platform\cite{Mikolajick2021}. 
In this regard, ferroelectric materials, which uniquely feature a spontaneous electric polarization switchable by an external electric field, are particularly attractive. 
When prepared in the form of highly-crystalline epitaxial thin films, it is further possible to finely tune the ferroelectric domain configuration, the direction and magnitude of the polarization, and the resulting macroscopic electrical properties of ferroelectric oxides, utilizing the epitaxial strain imposed by single-crystalline substrates\cite{Choi2004,Damodaran2016, Everhardt2016,Sarott2020,Muller2023}.
In CMOS, due to the substrate being restricted to silicon, the use of strain engineering is not applicable in general, which obstructs the integration of optimized ferroelectric thin films into advanced CMOS-based architectures.
Hence, rendering complex oxides, especially ferroelectrics, compatible with silicon has remained a major challenge. 

Among the various ferroelectrics, the perovskite-structure material BaTiO\textsubscript{3} (BTO) has been at the forefront of CMOS integration due to its robust ferroelectric properties in the thin-film regime, large dielectric constant, and absence of volatile or toxic elements. 
Several attempts have been made to integrate BTO on silicon substrates for diverse applications, including non-volatile memory\cite{Niu2011}, ferroelectric tunnel junctions (FTJs)\cite{Guo2015}, and electro-optic devices\cite{Abel2019, Xiong2014, Wen2024}. 
For these applications, the fabrication of BTO thin films on silicon has been accomplished using a number of deposition methods, including atomic layer deposition (ALD)\cite{Ngo2014}, molecular beam epitaxy (MBE)\cite{Mazet2015, Niu2007}, radio-frequency magnetron sputtering\cite{Wague2020}, and pulsed laser deposition (PLD)\cite{Chen2022, Bagul2025, Jiang2022, Silva2013, Wang2020}. 
A common difficulty that persists when attempting to grow high-quality BTO, as well as other complex oxides, on silicon, independent of the deposition method, however, is the large structural mismatch between the oxide and silicon. 
To address this issue, the use of buffer layers, such as SrTiO\textsubscript{3} (STO)\cite{Jiang2022, Ngo2014}, yttria-stabilized zirconia\cite{Lyu2018_SciRep}, and other perovskites\cite{Bagul2025} has been a popular strategy to both enable a controlled growth of BTO with varying thickness and tune the ferroelectric properties via strain engineering\cite{Lyu2018_SciRep}. 
This, hence, showcases that lattice matching using suitable buffer layers might be an auspicious route to optimize the properties of BTO thin films on silicon.

Despite many promising recent developments, even with the help buffer layers, the large mismatch in the coefficients of thermal expansion between Si and oxides typically introduces significant \emph{thermal strain}\cite{Warusawithana2009,Baek2013,Chen2022} upon cooling from the elevated growth temperatures of oxides, which can lead to extended structural defects. 
In ferroelectrics, this thermal strain can further trigger the uncontrolled formation of domains\cite{Dubourdieu2013}, which, in turn, can detrimentally impact the macroscopic electric-field response. 
Furthermore, for BTO thin films on silicon due to the tensile nature of the thermal strain, it has hitherto remained difficult to maintain a large remanent out-of-plane polarization with a low coercive field ($<$\SI{1}{\volt}), while simultaneously minimizing imprint. 
Imprint refers to a built-in bias in a ferroelectric heterostructure that leads to a preferential direction of a polarization and manifests as a horizontal shift in the hysteresis loop\cite{Zhou2005}. 
In extreme cases, imprint can entirely destabilize one polarization direction in the absence of an electric field (i.e.\ at remanence) and render a ferroelectric unipolar, such that it can only be switched in a volatile manner. 
Generally, imprint can originate from multiple sources with the most common one in epitaxial thin films being the presence of asymmetric top and bottom electrodes with different work functions\cite{Karthik2012, Do2020}.
Additionally, inhomogeneous epitaxial strain, defect dipoles, and surface defects\cite{Zhou2005} can also significantly contribute to imprint and hamper the realization of reliable and truly nonvolatile memory devices\cite{Zhou2005}.

In this work, we present imprint-free BTO thin films on silicon substrates with a low coercivity, a high remanent polarization, and an excellent fatigue resilience ($>10^{10}$ cycles). 
We achieve this by introducing a strain-mediating layer of SrSn\textsubscript{1-x}Ti\textsubscript{x}O\textsubscript{3} (SSTO) acting as a pseudo-substrate on STO-buffered Si substrates. 
By purposefully relaxing this SSTO layer, we minimize the thermal strain experienced by the BTO layer and simultaneously mimic on silicon the moderate compressive epitaxial constraints of the commonly used scandate substrate material GdScO\textsubscript{3} (GSO). 
Remarkably, our BTO films grown by pulsed laser deposition exhibit extremely low leakage, enabling the fabrication of ultrathin BTO devices with robust ferroelectric properties. 
In conjunction with good polarization retention and high fatigue resistance, this work thus paves the way for Si-compatible ferroelectric devices such as ferroelectric field-effect transistors or FTJs for low-energy consuming oxide electronics.

\section{Results and discussion}
\begin{figure}
    \centering
    \includegraphics[width=\linewidth]{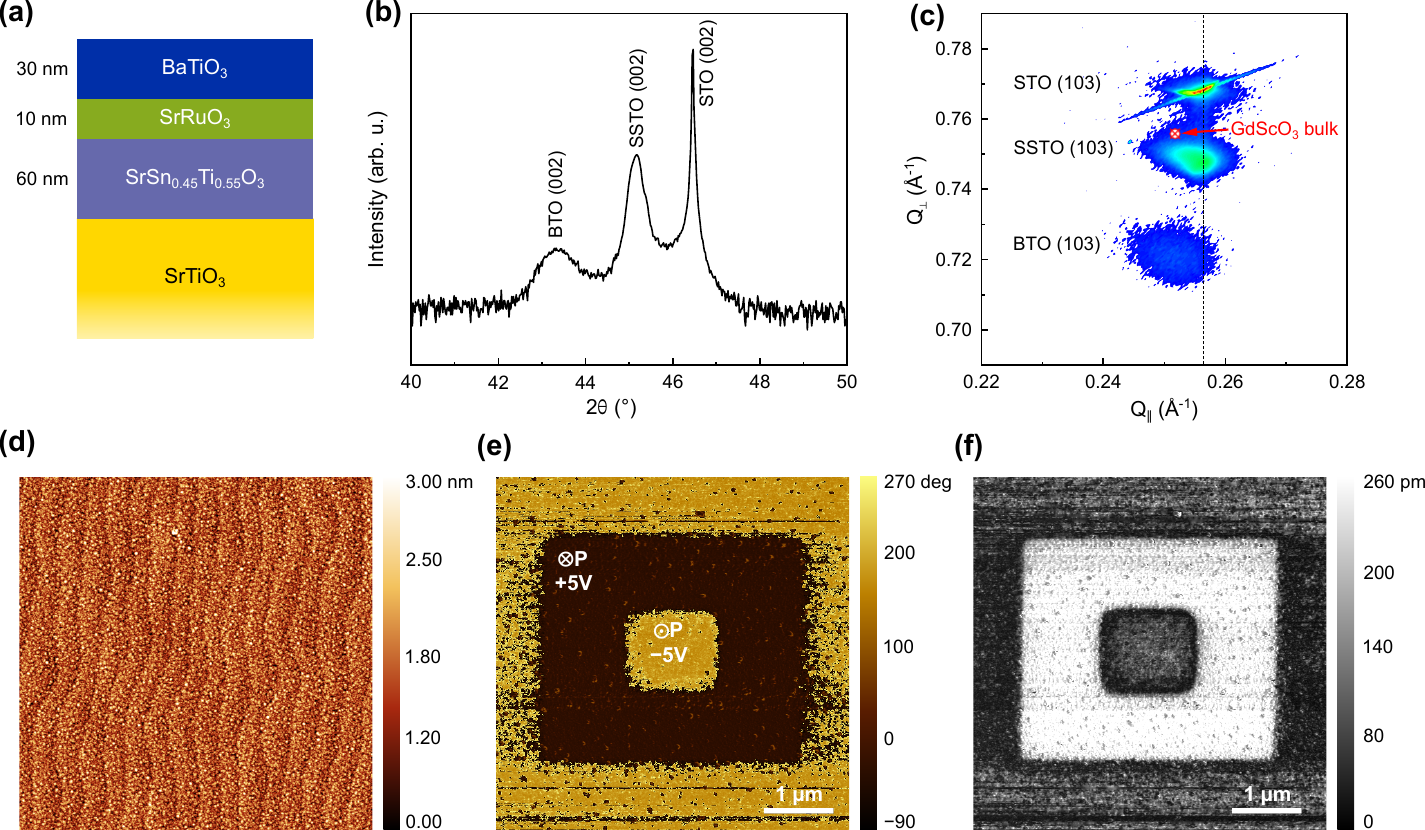}
    \caption{\textbf{Structural and local ferroelectric characterization of BaTiO\textsubscript{3} (BTO) on SrSn\textsubscript{0.45}Ti\textsubscript{0.55}O\textsubscript{3} (SSTO)-buffered SrTiO\textsubscript{3} (STO).} a) Illustration of the heterostructure BTO/SRO/SSTO grown on TiO\textsubscript{2}-terminated STO. b) Symmetric  XRD $\theta\text{--}2\theta$ scan of the heterostructure around the STO 002 peak. c) XRD reciprocal space map around the STO 103 reflection. d) \SI{5}{}$\times$\SI{5}{\micro\meter\squared} atomic force microscopy topographic image. Vertical piezoresponse force microscopy (PFM) phase e) and amplitude f) responses across an electrically poled box-in-box region.}
    \label{fig:1}
\end{figure}

We start our investigation by optimizing the growth of the BTO-based heterostructure on TiO\textsubscript{2}-terminated STO (see Methods), before moving to STO-buffered Si substrates. 
To do so, we first deposit a \SIadj{60}{\nano\meter}-thick layer of SrSn\textsubscript{0.45}Ti\textsubscript{0.55}O\textsubscript{3} (hereafter referred to as SSTO), followed by a \SIadj{10}{\nano\meter}-thick SrRuO\textsubscript{3} (SRO) bottom electrode, and a \SI{30}{\nano\meter} BTO layer (see Fig.\ \ref{fig:1}a). 
For all layers, we monitor the film thickness \emph{in-situ} with unit-cell precision using reflection high-energy electron diffraction (RHEED) and confirm it \emph{ex-situ} via X-ray reflectivity.
By tailoring the B-site cation ratio in SSTO, i.e.\ the Sn:Ti ratio, it is possible to linearly tune its lattice parameter to any value between \SI{4.033}{\angstrom} (pure SrSnO\textsubscript{3}) and \SI{3.905}{\angstrom} (pure SrTiO\textsubscript{3})\cite{Liu2015,Hamming-Green}.
Here, in order to emulate the epitaxial constraints of GdScO\textsubscript{3} ($a_{pc}=$\SI{3.963}{\angstrom}) with the SSTO pseudo-substrate layer, we select an SSTO composition of SrSn\textsubscript{0.45}Ti\textsubscript{0.55}O\textsubscript{3}. 
BTO films grown under moderate compressive strain on GdScO\textsubscript{3} (\SI{-0.9}{\percent}) have been shown to exhibit a robust out-of-plane polarization that persists up to large film thicknesses without the formation of in-plane polarized \emph{a}-domains\cite{Schubert2003,Strkalj2019,Jiang2022}.
Hence, by purposefully letting the SSTO layer relax on STO, it should be possible to obtain BTO-based heterostructures with ideal elastic boundary conditions for out-of-plane ferroelectricity on silicon, while preserving (quasi-)epitaxial quality. 

To characterize the out-of-plane crystallographic orientation of the epitaxial heterostructures, symmetric $\theta\text{--}2\theta$ X-ray diffraction (XRD) measurements are performed, as shown in Fig.\ \ref{fig:1}b.
Distinct 002 reflections from the BTO, SSTO, and STO layers are observed, indicating that all layers are well-aligned along the [001] direction. 
The absence of secondary phases or polycrystalline peaks confirms the high crystallinity of the films and, notably, we do not observe any evidence for \emph{a}-domain formation that would give rise to BTO 200/020 reflections. 
The reciprocal space map (RSM) around the STO 103 peak, shown in Fig.\ \ref{fig:1}c, further reveals the epitaxial relationship between the layers in the heterostructure. 
Specifically, we observe that both the SSTO and the BTO layers are partially relaxed from their respective underlayer, i.e. the STO and the SSTO, respectively. 
The atomic force microscopy (AFM) image in Fig.\ \ref{fig:1}d, shows that the BTO surface is atomically flat without detectable islands ($R_q<$\SI{4}{\angstrom}) and clear step-edge terraces, pointing toward ideal two-dimensional layer-by-layer growth. 
The ferroelectric nature and local switching behavior of the BTO film is further confirmed by piezoresponse force microscopy (PFM). 
Fig.\ \ref{fig:1}e and Fig.\ \ref{fig:1}f display the vertical PFM phase and amplitude images, respectively, recorded after applying electrical biases of $\pm$\SI{5}{\volt} to the PFM tip in a box-in-box pattern. 
The phase image clearly exhibits a \SI{180}{\degree} reversal of polarization upon the application of opposite biases, demonstrating the capability to controllably write, read, and reversibly switch ferroelectric domains. 
Furthermore, the amplitude exhibits a clear minimum at the location of the \SI{180}{\degree} domain walls (Fig.\ S1). 
In brief, the introduction of the SSTO buffer layer as a pseudo-substrate effectively reduces the lattice mismatch of the BTO film with the STO substrate, resulting in a purely out-of-plane polarized BTO film that is reversibly switchable.


\begin{figure}
    \centering
    \includegraphics[width=\linewidth]{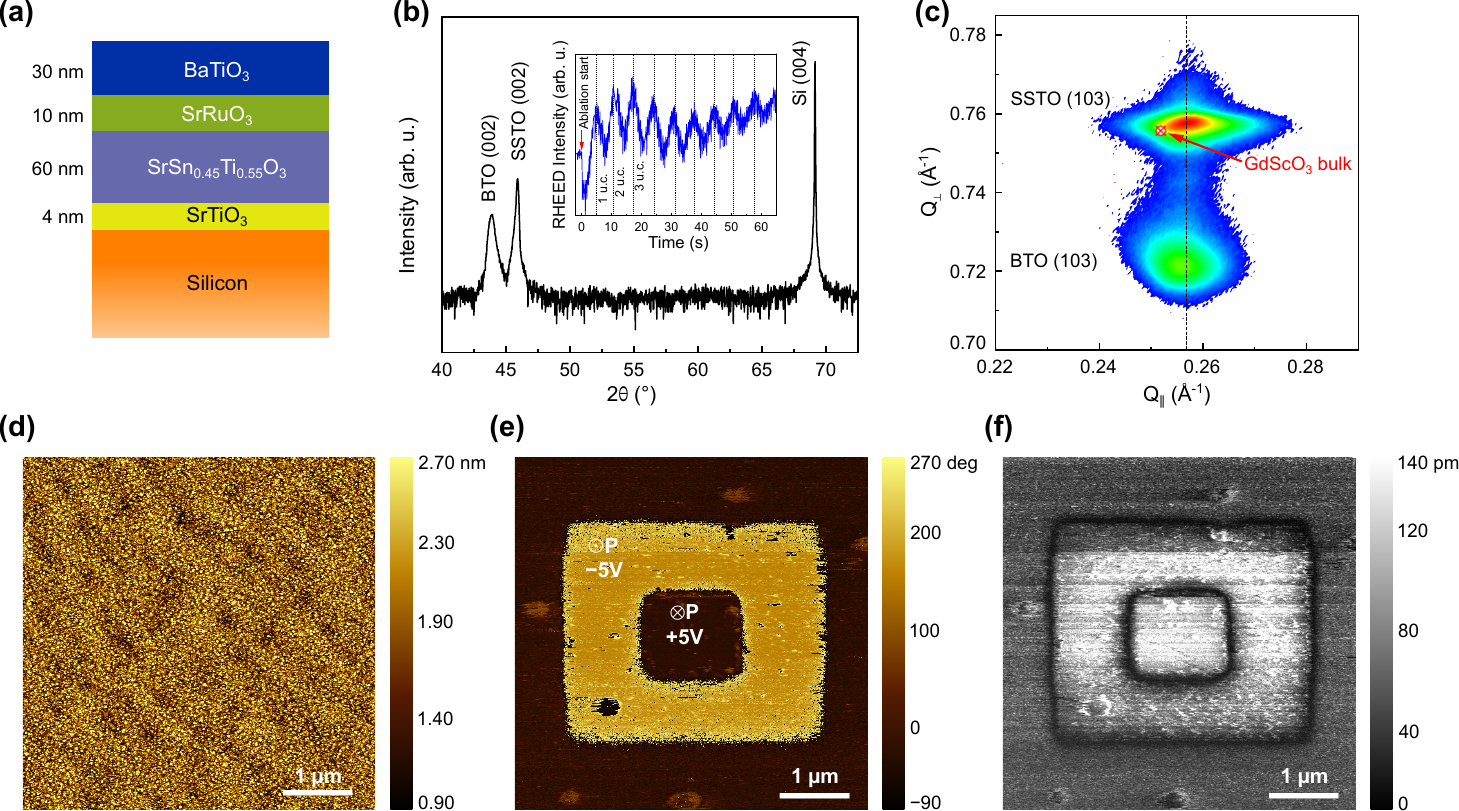}
    \caption{\textbf{Structural and local ferroelectric characterization of BTO on Si.} a) Illustration of the BTO/SRO/SSTO heterostructure grown on STO-buffered Si (001). b) Symmetric XRD $\theta\text{--}2\theta$ scan of the heterostructure with the inset showing the \emph{in-situ} acquired integrated RHEED time trace during BTO growth. c) XRD reciprocal space map around the SSTO 103 reflection, showing that the SSTO lattice parameters are close to those of GdScO\textsubscript{3} (known to provide optimal strain conditions for out-of-plane polarized BTO). d) \SI{5}{}$\times$\SI{5}{\micro\meter\squared} atomic force microscopy topographic image of the BTO surface. Vertical piezoresponse force microscopy phase e) and amplitude f) responses across the electrically poled box-in-box region.}
    \label{fig:2}
\end{figure}

Next, to check whether this buffer-layer strategy indeed enables us to maintain the high structural quality and the single-domain configuration of the BTO film on Si, we prepare a series of BTO films with varying thicknesses by PLD on STO-buffered Si $(001)$ substrates (see Figs.\ \ref{fig:2}a and S2). 
For simplicity, we focus our discussion here on a representative sample with a BTO thickness of \SI{30}{\nano\meter}, while the thicknesses of the underlying SSTO and SRO layers are kept constant at \SI{60}{\nano\meter} and \SI{10}{\nano\meter}, respectively.
The specular XRD $\theta\text{--}2\theta$ scan of this heterostructure, shown in Fig.\ \ref{fig:2}b, shows clear 00l reflections of BTO and SSTO, and confirm the preservation of the c-axis-oriented growth of BTO on silicon (see also Fig.\ S3). 
Again, no secondary phases, such as \emph{a}-domains, are detected and only the pure perovskite phase with a tetragonal structure is observed, analogously to the BTO-based heterostructure grown on single-crystalline STO in Fig.\ \ref{fig:1}. 
Furthermore, as shown in the inset of Fig.\ \ref{fig:2}b, during BTO growth, we observe pronounced RHEED intensity oscillations, which indicate near-ideal layer-by-layer growth.
While not being essential for achieving high-quality BTO films on STO-buffered Si substrates\cite{Jiang2022}, having \emph{in-situ} thickness control with unit-cell precision is important for the realization of ultrathin ferroelectric tunneling-based devices and confirms the optimization of the growth parameters.

In contrast to the BTO grown on bulk STO, the RSM around the SSTO 103 reflection on STO-buffered Si, shown in Fig.\ \ref{fig:2}c, reveals a fully relaxed state of the SSTO with its lattice constants lying very close to that of bulk GdScO\textsubscript{3}, whereas the BTO film becomes fully strained to the underlying SSTO. 
The complete strain relaxation of the SSTO in this case can be likely attributed to the compressive strain state of the thin STO buffer layer on Si $(001)$ compared to bulk STO\cite{Choi2012}, which increases the lattice mismatch between STO and SSTO and promotes the relaxation of the latter.
Hence, despite using different substrates, the $\theta\text{--}2\theta$ and RSM measurements in Figs.\ \ref{fig:1} and \ref{fig:2} showcase that the BTO/SRO/SSTO heterostructures maintain the same structure and out-of-plane orientation. 
We attribute this to the role of the SSTO layer, which successfully mimics the epitaxial constraints of the GdScO\textsubscript{3} substrate and, therefore, effectively decouples the BTO layer from the mechanical boundary conditions of the substrate.

The AFM image of the BTO surface in Fig.\ \ref{fig:2}d further shows an atomically-flat surface morphology comparable to that of the sample grown on bulk STO and consistent with the two-dimensional growth mode evidenced by RHEED. 
The vertical PFM phase (Fig.\ \ref{fig:2}e) and amplitude (Fig.\ \ref{fig:2}f) images after DC poling using the PFM tip reveal reversible out-of-plane ferroelectric switching, comparable to that of BTO on STO. 
Notably, the pristine out-of-plane orientation of the BTO polarization on STO-buffered Si is predominantly downward and, thus, opposite to the upward-oriented polarization observed in the BTO films grown on STO. 
This difference may originate either from variations in substrate/interface termination, defect chemistry, or the minor difference in the epitaxial boundary conditions\cite{Yu2012,Weymann2020,Sarott2023, Sarott2020}.
Importantly, in comparison to previous studies that have reported strong depolarization effects in BTO films on STO-buffered Si, which cause spontaneous back-switching of poled domains within \SI{10}{\minute}\cite{Jiang2022, Dubourdieu2013}, our films maintain a clear PFM contrast for at least \SI{100}{\minute} (see Fig.\ S4). 
This enhanced polarization stability underscores the improved structural and chemical homogeneity of our BTO films via a reduction of thermal stresses and improved lattice matching enabled by the SSTO buffer.


\begin{figure}
    \centering
    \includegraphics[width=\linewidth]{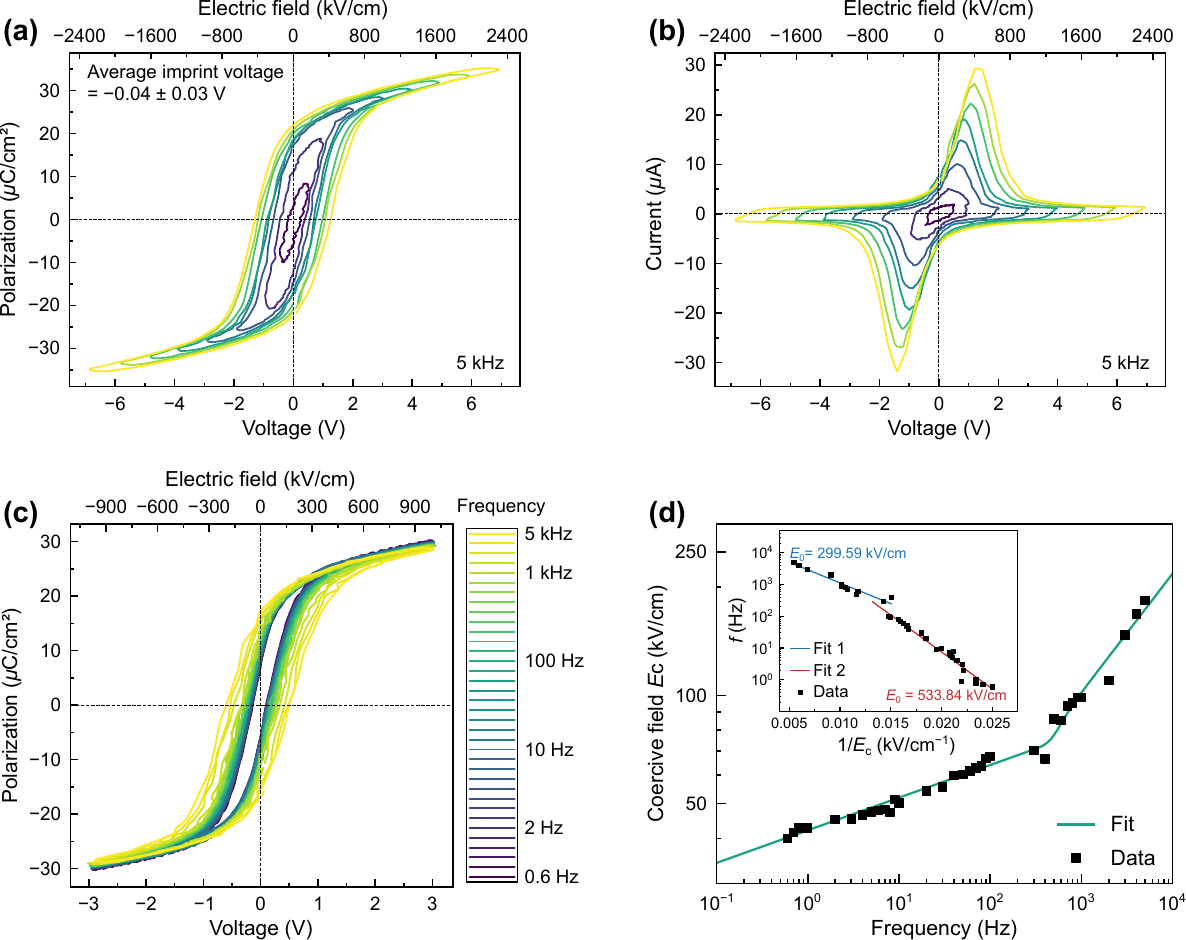}
    \caption{\textbf{Macroscopic ferroelectric characterization of BTO on Si.} a) Polarization-electric field hysteresis loops  measured at \SI{5}{\kilo\hertz} for various electric field amplitudes, showing lack of imprint. b) Switching current profiles corresponding to the loops in a). c) Frequency-dependent hysteresis loops over the range \SI{0.6}{\hertz}--\SI{5}{\kilo\hertz} showing coercive voltages ($V_\mathrm{c}$) in the range of $0.12-$\SI{0.55}{\volt}. d) Coercive field ($E_\mathrm{c}$) as a function of frequency extracted from c). The inset shows the Merz-law fitting of the same dataset.}  
    \label{fig:3}
\end{figure}

Realistic ferroelectrics-based devices require controlled and repeatable polarization switching in a parallel-plate geometry. 
To investigate the macroscopic ferroelectric properties of our BTO heterostructures, we therefore fabricate BTO capacitor structures with symmetric SRO electrodes (i.e.\ SRO/BTO/SRO/SSTO on STO-Si). 
Fig.\ \ref{fig:3}a shows the $P$-$E$ hysteresis loops of a \SIadj{30}{\nano\meter}-thick BTO capacitor, measured at a frequency of \SI{5}{\kilo\hertz} across a range of voltages.
Upon increasing voltage amplitude, the $P$-$E$ loops become well-saturated with remanent polarization ($P_\mathrm{r}$) values of approximately \SI{20}{\micro\coulomb\per\square\centi\meter}, which closely aligns with the values reported for bulk BTO and comparable to those observed in BTO thin films on oxide substrates\cite{Trithaveesak2005}. 
Most notably, in comparison to previous reports of BTO films on Si\cite{Jiang2022,Bagul2025}, the BTO films here exhibit a lower coercive field value of \SI{0.61}{\volt} (\SI{203.76}{\kilo\volt\per\centi\meter}), while being essentially free of imprint $\frac{V_\mathrm{c,+}+V_\mathrm{c,-}}{2}=0.04\pm \SI{0.03}{\volt}$. 
The absence of imprint indicates that the insertion of the SSTO buffer is highly effective against the formation of built-in fields that can pin the ferroelectric polarization. 
This includes previously reported structural causes of imprint, such as strain gradients due to thermal strain or oriented defect dipoles\cite{Damodaran2014, Scott2000} that are commonly observed even when employing symmetric electrodes. 
In addition to the lack of imprint, the corresponding current-electric field ($I$-$E$) profiles (Fig.\ \ref{fig:3}b) show a remarkable absence of leakage, even for applied electric fields ($>$\SI{2}{\mega\volt\per\centi\meter}) that by far exceed the coercivity.

To further deepen our understanding of the BTO switching dynamics, we investigate the frequency dependence of the hysteresis loops. Fig.\ \ref{fig:3}c contains the BTO $P$-$E$ loops over a frequency range from \SI{0.6}{\hertz} to \SI{5}{\kilo\hertz} with a fixed maximum applied voltage of \SI{3}{\volt}. 
For all measured frequencies, the BTO films remain free of leakage and imprint, but we observe a clear reduction of the coercive voltage with decreasing frequency from \SI{0.55}{\volt} at \SI{5}{\kilo\hertz} to \SI{0.12}{\volt} at \SI{0.6}{\hertz}.
Such a frequency dependence of the coercive field is, in fact, consistent with reports in many ferroelectric bulk crystals and thin films, where higher frequencies are typically accompanied by a notable increase in the coercive field\cite{Yang2010, Chen2017, Liu2006, Scott1996}.
Within the theoretical Ishibashi-Orihara framework based on the Avrami model\cite{Ishibashi1995, Orihara1994, Chen2017}, the frequency dependence of the coercive field follows a power law $E_\mathrm{c} \propto f^{\beta}$, implying a linear relation in a $\log(E_\mathrm{c})$ vs.\ $\log(f)$ plot. 
For the BTO films considered here, however, the $\log(E_\mathrm{c})$ vs.\ $\log(f)$ plot, shown in Fig.\ \ref{fig:3}d, clearly exhibits two distinct linear scaling regimes with a crossover at around \SI{500}{\hertz} that cannot be fitted with a single power-law exponent $\beta$.
This, hence, implies two distinct types of switching kinetics contributing to polarization reversal, as observed similarly for bulk relaxor PMN-PT crystals\cite{Chen2017} and PZT thin films\cite{Yang2010, Karthik2012,Liu2006}. 
Taking into account two distinct switching kinetics regimes and using the empirical expression proposed by Chen \emph{et al.}\ \cite{Chen2017}: 
\begin{equation}
E_\mathrm{c} = K_1 f^{\beta_1} \left[ 1 - \exp\!\left( -\frac{1}{\tau_1\cdot f} \right) \right]
+ K_2 f^{\beta_2} \exp\!\left( -\frac{1}{\tau_2\cdot f} \right)
\label{eq:empirical}
\end{equation}
where $\tau_1$,$\tau_2$ correspond to the characteristic relaxation times associated with domain reorientation, we are indeed able to fit our data. 
Note that the obtained values for $\tau_{1} = \SI{4.07}{\milli\second}$ and $\tau_{2} = \SI{11.3}{\milli\second}$ are of the same order of magnitude as the inverse of the crossover frequency at \SI{500}{\hertz}, which is consistent with the trend described by Chen \emph{et al.}\cite{Chen2017}.

Interestingly, our measured crossover frequency of $\sim$\SI{500}{\hertz} lies close to that of $200-$\SI{500}{\hertz} reported for PZT thin films \cite{Yang2010}. 
This similarity points toward a common underlying mechanism for the observed change in domain-switching dynamics with thermally-activated domain-wall creep dominating polarization switching at low frequencies and viscous domain-wall flow governing at high frequencies\cite{Yang2010,Tuckmantel2021}.

Subsequently, applying Merz' law \cite{Merz1954, Shin2007} to the frequency dependence of $E_\mathrm{c}$
\begin{equation}
f = f_{0} \exp\left( -\frac{E_{0}}{E_\mathrm{c}} \right)
\label{eq:merz}
\end{equation}
allows us to quantify the activation fields $E_{0}$ of the high- and low-frequency switching regimes.
To do so, we plot $\log(f)$ vs.\ $(\num{1}/E_\mathrm{c})$, as shown in the inset of Fig.\ \ref{fig:3}d, where we extract activation fields of $\SI{533.8}{\kilo\volt\per\centi\meter}$ in the low-frequency and $\SI{299.6}{\kilo\volt\per\centi\meter}$ in the high-frequency regime.
For comparison, similar measurements in PMN--PT single crystals have yielded activation field values ranging between $20-$\SI{70}{\kilo\volt\per\centi\meter}\cite{Chen2017}, whereas in epitaxial oxide thin films activation fields have been found to lie broadly spread between \SI{100}{\kilo\volt\per\centi\meter} and $>$\SI{1000}{\kilo\volt\per\centi\meter}\cite{Chen1999}.


\begin{figure}
    \centering
    \includegraphics[width=\linewidth]{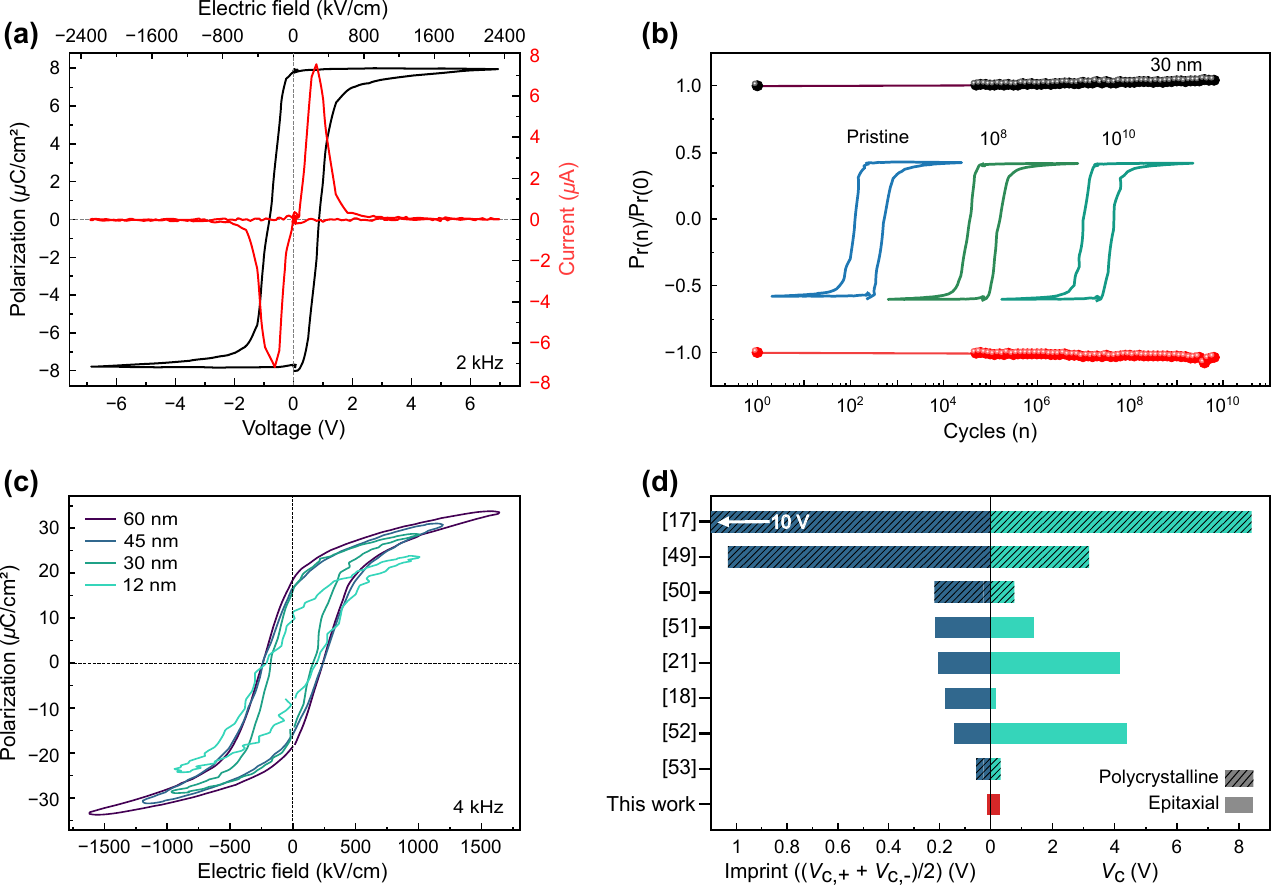}
    \caption{\textbf{Robustness of the BTO switching characteristics.} a) PUND hysteresis loops and corresponding switching current profiles measured at \SI{2}{\kilo\hertz}. b) Ferroelectric cycling endurance over $10^{10}$ cycles measured with a voltage amplitude of \SI{4}{\volt} and a frequency of \SI{2}{\kilo\hertz}. The inset show the PUND loops obtained after 1, $10^{8}$, and $10^{10}$ cycles. c) Dynamic hysteresis loops for BTO films on Si with different thickness. d) Literature comparison of the device imprint and $V_\mathrm{c}$ in BTO thin films on Si substrates, with the vertical axis indicating the reference numbers.}  
    \label{fig:4}
\end{figure}

Dynamic hysteresis measurements, such as those in Fig.\ \ref{fig:3}, contain multiple current contributions from ferroelectric switching, capacitive charging, or leakage, and therefore often lead to an overestimation of the intrinsic ferroelectric polarization.
In addition, they fail to capture spontaneous back-switching due to depolarization.
To assess the true intrinsic remanent polarization of our films, we perform positive-up-negative-down (PUND) measurements, as shown in Fig.\ \ref{fig:4}a, for a \SIadj{30}{\nano\meter}-thick BTO capacitor. 
The obtained 2$P_\mathrm{r}$ value of \SI{16}{\micro\coulomb\per\square\centi\meter} indicates that our BTO films on Si exhibit a strong intrinsic polarization after eliminating all non-ferroelectric current contributions.
This polarization further remains robust under electric field cycling without any noticeable fatigue even after $10^{10}$ cycles, as evident from Fig.\ \ref{fig:4}b.
Note that in comparison to previous studies of BTO on Si\cite{Jiang2022}, our fatigue measurements do not require biasing to compensate for device imprint.
Hence, by getting rid of imprint in our BTO heterostructures via the introduction of the  strain-mediating SSTO pseudo-substrate, we essentially eliminate fatigue, one of the most serious reliability concerns in ferroelectric films, and surpass previously reported BTO cycling endurance limits.
This likely points to a superior interfacial quality and a suppressed influence of intrinsic defects, such as oxygen vacancies, on the cycling performance.

The robustness of the ferroelectric switching characteristics is also maintained when varying the thickness of the BTO layer in the heterostructures. 
The dynamic hysteresis measurements in Fig.\ \ref{fig:4}c display saturated, symmetric, and imprint-free ferroelectric switching characteristics even when reducing the BTO thickness to \SI{12}{\nano\meter} (see also Figs. S5, S6, and S7). 
The structural properties of the BTO layer also remain unchanged over the full thickness range (see Fig.\ S2). 
Furthermore, as summarized in Fig.\ \ref{fig:4}d and Tab.\ S1, a literature comparison of the imprint in BTO-based thin films on Si substrates shows that the imprint of our epitaxial BTO thin films is substantially lower than previously reported values\cite{Dwivedi2018, Bagul2025, Qiao2010, Scigaj2013, Lyu2018_SciRep, Jiang2022, Yamada2016, Drezner2003}.
Achieving imprint-free ferroelectric devices on silicon constitutes a key step toward reliable long-term cycling and high-frequency operation, and promotes their integration into advanced ferroelectric memory and logic devices.

\subsection{Conclusion}
In short, we demonstrate the epitaxial integration of BTO thin films on STO-buffered silicon using a SSTO buffer layer that acts as a strain-mediating pseudo-substrate and eliminates any parasitic structural degradation due to pervasive thermal strain.
The prepared BTO thin films are highly crystalline with a pure out-of-plane orientation, exhibiting low coercivity, absence of imprint, and exceptional fatigue resistance ($>$$10^{10}$ cycles).
Thus, our study enables reliable ferroelectric device operation with large potential benefits for BTO-based CMOS-compatible energy-efficient memory and logic applications. 

\section{Experimental}

\textbf{Sample preparation and structural characterization}.\\
All thin films in this study are deposited by pulsed laser deposition using a \SI{248}{\nano\meter} KrF excimer laser on $(001)$-oriented STO substrates (CrysTec GmbH) and STO-buffered Si substrates (Lumiphase AG). Prior to the deposition process, the STO $(001)$ substrates are treated with buffered hydrofluoric acid and annealed to produce a TiO\textsubscript{2}-termination surface exhibiting atomically smooth terraces. 
The SSTO buffer layers are grown at \SI{650}{\degreeCelsius} under an oxygen partial pressure of \SI{0.3}{\milli\bar} using a laser fluence of \SI{1.43}{\joule\per\square\centi\meter} at \SI{2}{\hertz}. The bottom electrode SRO is then deposited at \SI{650}{\degreeCelsius} with an oxygen partial pressure of \SI{0.015}{\milli\bar} at the same laser fluence and repetition rate. BTO thin films with varying thicknesses ($12-$\SI{60}{\nano\meter}) are subsequently grown at \SI{650}{\degreeCelsius} under an oxygen partial pressure of \SI{0.015}{\milli\bar} at a laser fluence of \SI{1.48}{\joule\per\square\centi\meter} at \SI{2}{\hertz}. Finally, the top electrode SRO is deposited under the identical growth conditions with the bottom SRO layer. All layers are grown at a target-substrate distance of \SI{50}{\milli\meter}.  Film thicknesses are tracked in-situ during growth by reflection high-energy electron diffraction and confirmed by ex-situ X-ray reflectivity measurements.
\newline
$\theta\text{--}2\theta$ scans and reciprocal space maps are carried out on a Panalytical X'Pert MRD thin-film diffractometer using a Cu source (CuK$\alpha$, \SI{1.540598}{\angstrom}) with a 2xGe(220) hybrid monochromator and a PIXcel\textsuperscript{3D} area detector.\\ 

\noindent \textbf{Devices Fabrication and Electrical Characterization}.\\
All electrical measurements are performed on circular capacitor structures. Top SRO electrode pads with different sizes are fabricated by photolithography (Karl Suss mask aligner MA1006) and ion beam etching (Intlvac Nanoquest Pico~4RF). The dynamic $P$-$E$ loops, $I$-$E$ curves, PUND measurement, and fatigue measurements are performed on a ferroelectric analyzer (TF analyzer~2000, aixACCT). To investigate the frequency dependence of $P$-$E$ loops, triangular waves with frequencies ranging from \SI{0.6}{\hertz} to \SI{5}{\kilo\hertz} are applied.\\

\noindent \textbf{AFM and PFM Characterization}.\\
Atomic force microscopy measurements are performed in tapping mode on a Bruker Dimension Icon scanning probe microscope equipped with Tap300Al-G tips (Budget Sensors, $k=$ \SI{40}{\newton\per\meter}).
Piezoresponse force microscopy measurements are carried out on an Asylum Research Cypher ES scanning probe microscope equipped with PtIr-coated tips (Bruker SCM-PIT-V2, $k=$ \SI{3}{\newton\per\meter}). PFM measurements are done in air and on-resonance with a tip-sample force of approximately \qty{15}{\nano\newton}.\\

\section{Author contributions}
J.Z. and M.F.S. conceived and designed the experiments. J.Z. fabricated the films, conducted XRD measurements, electrical measurements, and data analysis with assistance from M.F.S. M.F.S. performed the PFM measurements. J.Z., M.F.S. and B.N. wrote the manuscript. M.F.S. supervised the work jointly with B.N. All authors discussed the results and commented on the manuscript.

\section{Competing financial interests}
The authors declare no conflict of interest. 

\section{Data availability}
The data supporting this study are available in DataverseNL at\\ \href{https://doi.org/10.34894/YMBX4L}{https://doi.org/10.34894/YMBX4L}.

\begin{acknowledgement}

The authors thank Jacob Baas, Henk Bonder and Joost Zoestbergen for technical support. J.Z., B.N., and M.F.S. acknowledge fincancial support from the Groningen Cognitive Systems and Materials Center (CogniGron) and the Ubbo Emmius Foundation of the University of Groningen. J.Z. acknowledges the support of the China Scholarship Council program (Project ID: 202206310041). B.N. and M.F.S. acknowledge funding from the HORIZON-CL4-2023 grant CONCEPT (101135946). M.F.S. acknowledges funding from the HORIZON-MSCA-2024-PF RECOMPUTE (101203197).

\end{acknowledgement}

\begin{suppinfo}

The Supporting Information contains additional structural and electrical measurements along with a table containing a literature comparison of the relevant ferroelectric metrics for BaTiO\textsubscript{3} thin films grown on silicon substrates.

\end{suppinfo}

\providecommand{\latin}[1]{#1}
\makeatletter
\providecommand{\doi}
  {\begingroup\let\do\@makeother\dospecials
  \catcode`\{=1 \catcode`\}=2 \doi@aux}
\providecommand{\doi@aux}[1]{\endgroup\texttt{#1}}
\makeatother
\providecommand*\mcitethebibliography{\thebibliography}
\csname @ifundefined\endcsname{endmcitethebibliography}  {\let\endmcitethebibliography\endthebibliography}{}

\end{document}